\documentstyle[preprint,aps,12pt,epsf]{revtex}
\tightenlines
\begin{document}
\draft
\preprint{\vtop{{\hbox{YITP-00-65}\vskip-0pt
}}}
\title{Non-factorizable contributions in $B$ decays revisited
}
\author{K. Terasaki\\ Yukawa Institute for Theoretical Physics,\\
Kyoto University, Kyoto 606-8502, Japan
}
\maketitle
\thispagestyle{empty}
\begin{abstract}

$\bar B \rightarrow D\pi$, $D^*\pi$, $J/\psi\bar K$ and $J/\psi\pi$ 
decays are studied by decomposing their amplitude into a sum of 
factorizable and non-factorizable ones. The former is estimated by
using the naive factorization while the latter is calculated by using 
a hard pion approximation in the infinite momentum frame. 

Although experiments provide only upper bounds of the branching 
ratios for the color suppressed $\bar B^0 \rightarrow D^0\pi^0$ and 
$D^{*0}\pi^0$ decays, consistency among the branching ratios for the 
$\bar B \rightarrow D\pi$ and $D^*\pi$ decays including the charged 
modes leads to phenomenologically allowed lower limits of 
${\cal B}(\bar B^0 \rightarrow D^0\pi^0)$ and 
${\cal B}(\bar B^0 \rightarrow D^{*0}\pi^0)$. 
Our result is compared with the phenomenologically estimated 
branching ratios as well as the measured ones. As the consequence, 
it is seen that the non-factorizable amplitude is rather small in the 
color favored $\bar B \rightarrow D\pi$ and $D^*\pi$ decays but can 
still efficiently interfere with the main (factorized) amplitude. 
In the color suppressed $\bar B \rightarrow J/\psi\bar K$ and
$J/\psi\pi$ decays, non-factorizable contribution is more important. 
A sum of the factorized and non-factorizable amplitudes can improve 
the result from the factorization, although it is sensitive to 
model dependent values of form factors involved. 

\end{abstract}

\vskip 0.5cm
\newpage
\section{Introduction}

The factorization (or vacuum insertion) prescription was first
proposed long time ago~\cite{Oneda-Wakasa} and, after the discovery 
of the standard model, it was revived~\cite{SVZ}, has been improved 
and applied extensively to hadronic weak decays of heavy 
mesons~\cite{BSW,NRSX}. The improved one has been supported by two 
independent arguments, i.e., one is the large $N_c$ (color degree of 
freedom) argument~\cite{Large-N} that the factorizable amplitude 
which is given by the leading terms in the large $N_c$ expansion 
dominates in hadronic weak decays and the other is the color 
transparency argument\cite{DG} that the factorization works well 
under a particular kinematical condition, i.e., a heavy quark decays 
into another heavy quark plus a pair of light quark and anti-quark 
which are emitted colinearly with sufficiently high energies, for 
example, as $b \rightarrow c\,+\, (\bar ud)_1$, where $(\bar ud)_1$ 
denotes a color singlet $(\bar ud)$ pair. In the former case, if the 
factorization  works well in $B$ decays, it will again work well in 
$K$ and charm decays since the large $N_c$ argument is independent of 
flavors. On the other hand, in the latter case, it cannot work well 
in $K$ and charm decays where the kinematics cannot satisfy the above 
condition, even if it works well in $B$ decays. 

We here consider two body decays of charm mesons to check if the 
large $N_c$ argument works well in  hadronic weak interactions. A
naive application of the factorization prescription to charm decays 
leads to the color suppression [suppression of color mismatched 
decays, $D^0 \rightarrow \bar K^0\pi^0$, $\bar K^{*0}\pi^0$, etc., 
described by $c \rightarrow (s\bar d)_1 \,+\, u$]. It means, for 
example, that the amplitude for the $D$ meson decays into isospin 
$I={1\over 2}$ $(\bar K\pi)$ final states is approximately cancelled 
by the one into $I={3\over 2}$ $(\bar K\pi)$ final states and hence 
the phases of these amplitudes are nearly equal to each other. 
Therefore the factorized amplitudes for two body decays of charm 
mesons should be approximately real except for the overall phase. 
However the measured rates for these decays are not compatible with 
the color suppression and the amplitudes for 
$D \rightarrow \bar K\pi$ and $\bar K^*\pi$ decays have large phase 
differences between the amplitudes for decays into the $I={1\over 2}$ 
and ${3\over 2}$ final states~\cite{BHP}. To get rid of this problem, 
the factorization has been implemented by taking account for final 
state interactions. However, amplitudes with final state interactions 
are given by non-leading terms in the large $N_c$ expansion as will be 
discussed later, so that the large $N_c$ argument does not work well 
in charm decays and hence also in $B$ decays since the large $N_c$
argument is independent of flavors. On the other hand, if the color 
transparency argument works well, the factorization will be a good 
approximation in the $\bar B \rightarrow D\pi$ and $D^*\pi$ decays 
while, in the $\bar B \rightarrow J/\psi\bar K$ and $J/\psi\pi$, 
the factorization cannot be a good approximation (in contrast with in 
the large $N_c$ argument) but non-factorizable long distance 
contribution can play a role.  

In this article, we study $\bar B \rightarrow D\pi$, $D^*\pi$, 
$J/\psi\bar K$ and $J/\psi\pi$ decays. In the next section, we will 
review briefly the effective weak Hamiltonian and our basic 
perspective. In Sec.~III, the $\bar B \rightarrow D\pi$ and $D^*\pi$ 
decays will be studied by decomposing their amplitude into a sum of 
factorizable and non-factorizable ones. The former will be estimated 
by using the naive factorization while the latter is calculated by 
using a hard pion approximation in the infinite momentum frame (IMF). 
We will estimate phenomenologically allowed branching ratios from 
the measured ones and compare our result with the estimated ones as 
well as the measured ones. As the consequence, it will be seen that 
the factorization works fairly well but non-factorizable 
contributions are still not negligible. In Sec.~IV, the color 
suppressed decays, 
$\bar B \rightarrow J/\psi\bar K$ and $J/\psi\pi$, 
will be investigated in the same way. A brief summary will be given 
in the final section. 

\section{Effective Weak Hamiltonian}

Before we study amplitudes for $B$ decays, we review briefly the 
$|\Delta B|=1$ effective weak Hamiltonian. Its main part is usually 
written in the form 
\begin{equation}
H_w = {G_F \over \sqrt{2}}V_{ud}V_{bc}\Bigl\{c_1O_1 
                                      + c_2O_2 \Bigr\} +  h.c.,  
                                                    \label{eq:HW}
\end{equation}
where $c_1$ and $c_2$ are the Wilson coefficients of the four quark 
operators, $O_1$ and $O_2$, respectively, given by products of color 
singlet left-handed currents, 
\begin{equation}
O_1 = :(\bar du)_{V-A}(\bar cb)_{V-A}: \quad {\rm and} \quad 
O_2 = :(\bar cu)_{V-A}(\bar db)_{V-A}: .          \label{eq:FQO}
\end{equation}
The renormalization scale ($\mu$) dependence of $c_i$ and $O_i$ is 
not explicitly described unless it is required. $V_{ij}$'s denote the 
CKM matrix elements~\cite{CKM} which are taken to be real since CP 
invariance is always assumed in this article. 

When we calculate factorizable amplitudes for the 
$\bar B\rightarrow D\pi$ and $D^*\pi$ decays later, we use, as usual, 
the so-called BSW Hamiltonian~\cite{BSW,NRSX} 
\begin{equation}
H_w^{\rm BSW} = {G_F \over \sqrt{2}}V_{ud}V_{cb}
       \Bigl\{a_1O_1 + a_2O_2 \Bigr\} + h.c.   \label{eq:HW-BSW}
\end{equation}
which can be obtained from Eq.(\ref{eq:HW}) by using the Fierz 
reordering, where the operators $O_1$ and $O_2$ in Eq.(\ref{eq:HW-BSW}) 
should be no longer Fierz reordered. The coefficients $a_1$ and $a_2$ 
are given by 
\begin{equation}
a_1 = c_1 + {c_2 \over N_c},\quad a_2 = c_2 + {c_1 \over N_c},
                                               \label{eq:coef-BSW}
\end{equation}
where $N_c$ is the color degree of freedom. When $H_w^{\rm BSW}$ is 
obtained, an extra term which is given by a color singlet sum of 
products of colored currents, 
\begin{equation}
\tilde H_w = {G_F \over \sqrt{2}}V_{ud}V_{cb}
\Bigl\{c_2\tilde O_1 + c_1\tilde O_2 \Bigr\} + h.c., 
                                          \label{eq:HW-Non-f}
\end{equation}
comes out, where 
\begin{equation}
\tilde O_1 = 2\sum_a:(\bar dt^au)_{V-A}(\bar ct^ab)_{V-A}: 
\quad {\rm and} \quad 
\tilde O_2 = 2\sum_a:(\bar ct^au)_{V-A}(\bar dt^ab)_{V-A}:         
                                             \label{eq:FQO-extra}
\end{equation}
with the generators $t^a$ of the color $SU_c(N_c)$ symmetry. 

To realize physical amplitudes for $B$ decays by matrix elements of 
$\tilde H_w$, soft gluon(s) should be exchanged between quarks which 
belong to different meson states or different currents. Therefore, 
amplitudes given by $\tilde H_w$ correspond to non-leading terms in 
the large $N_c$ expansion and are not factorizable, i.e., 
$\tilde H_w$ is responsible for non-factorizable amplitudes. 
The above can be described schematically in Fig.~1, where we have not 
explicitly shown soft gluon exchange(s) between quarks belonging to 
the same meson and the same current for 
\newpage
\vspace{5mm}
\begin{center}
\epsfxsize=0.75\hsize
\epsffile{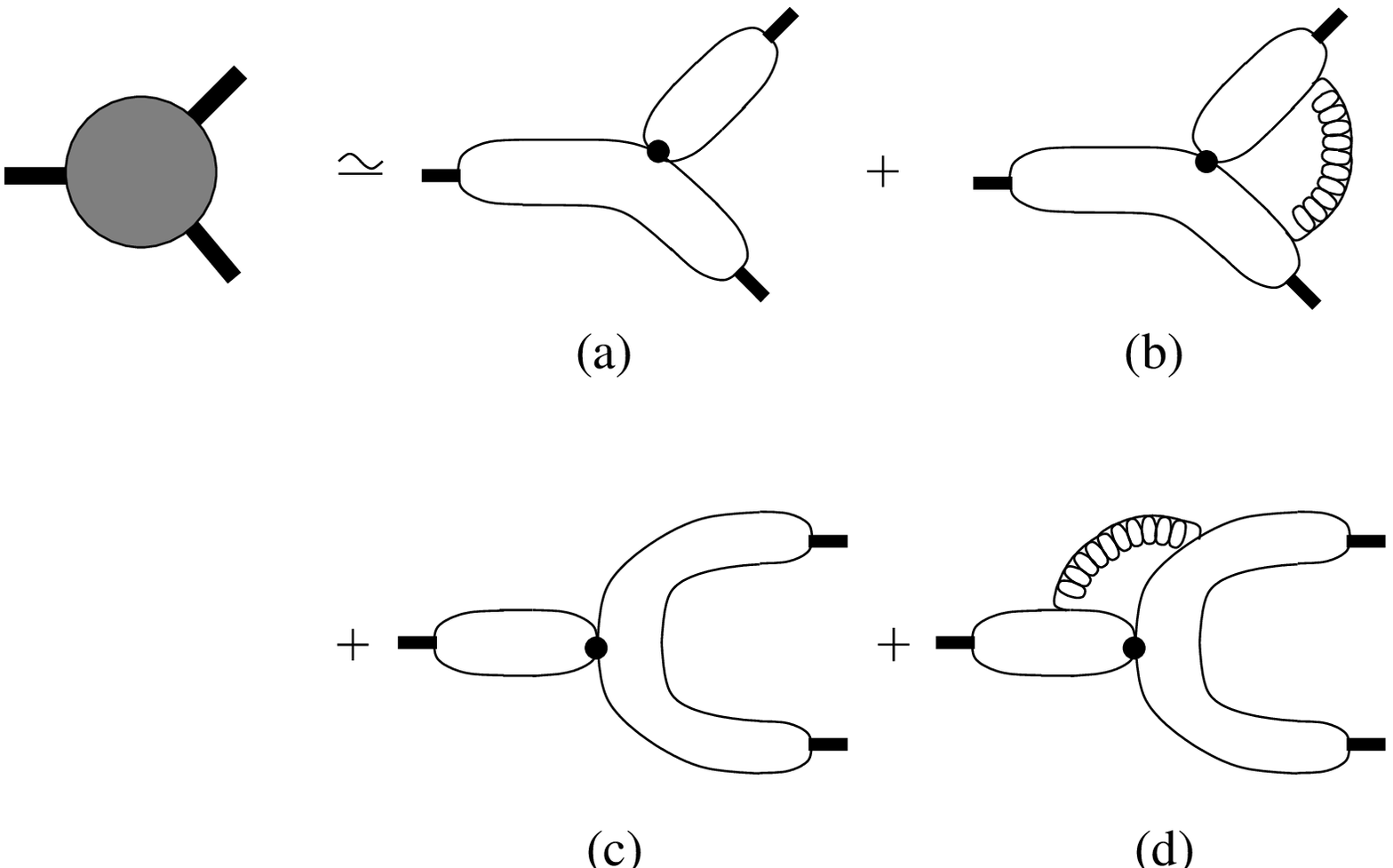}
\vspace{5mm}
\begin{minipage}{140mm}
{Fig.~1. Quark-line diagrams describing a hadronic weak decay of 
meson into two meson final states. Connectedness of quark-lines are
assumed. The bullet denotes the weak vertex with perturbative QCD
corrections and the spiral line denotes the soft-gluon(s). Soft-gluon
exchange(s) between quarks belonging to the same meson and the same
current are not explicitly described. The diagrams (a) and (c) 
correspond to the leading terms in the large $N_c$ expansion while 
(b) and (d) to the non-leading terms. 
}
\end{minipage}
\end{center}
\vspace{10mm}
simplicity. In the diagrams 
(a) and (c), no soft gluon(s) are exchanged between quarks belonging 
to different mesons or different currents. Therefore, the amplitudes 
described by these diagrams which correspond to the leading term in 
the large $N_c$ expansion are factorizable. In these diagrams, the 
weak vertices are given by products of color singlet currents as in 
$H_w^{\rm BSW}$. However, the diagrams (b) and (d) correspond to 
non-leading terms and are non-factorizable since soft gluon(s) are 
exchanged between quarks belonging to different mesons or different 
currents, so that the weak vertices in these diagrams are given by 
products of colored currents as in $\tilde H_w$. In the large $N_c$ 
argument, the latter is neglected while, in the color transparency, 
it is not guaranteed that the latter can be neglected. Therefore, 
$H_w^{\rm BSW}$ and $\tilde H_w$ are responsible for factorizable and 
non-factorizable amplitudes, respectively. 

\section{$\bar B \rightarrow D\pi$ and $D^*\pi$ decays}

As discussed in the previous section, we start to study nonleptonic 
weak decays decomposing their amplitude into a sum of 
{\it factorizable} and {\it non-factorizable} 
ones~\cite{Terasaki-B,TBD99}. The former is estimated by using the 
naive factorization in the BSW scheme~\cite{BSW}. The latter will be 
calculated later by using a hard pion approximation in the infinite 
momentum frame (IMF)~\cite{hard pion,suppl}. The hard pion amplitude 
will be given by {\it asymptotic} matrix elements of $\tilde H_w$ 
(matrix elements of $\tilde H_w$ taken between single hadron states 
with infinite momentum). 

Now we consider, as an example, the factorizable amplitude for the 
$B^-(p) \rightarrow D^0(p')\pi^-(q)$ decay. It is given by
\begin{eqnarray}
&& M_{\rm FA}(B^-(p) \rightarrow D^0(p')\pi^-(q))   \nonumber\\
&&\hspace{20mm}
= {G_F \over \sqrt{2}}V_{cb}V_{ud}
\Bigl\{ 
a_1\langle \pi^-(q)|(\bar du)_{V-A}|0\rangle 
\langle D^0(p')|(\bar cb)_{V-A}|B^-(p) \rangle      \nonumber\\   
&&\hspace{45mm} 
+ a_2\langle D^0(p')|(\bar cu)_{V-A}|0\rangle
\langle \pi^-(q)|(\bar db)_{V-A}|B^-(p) \rangle
\Bigr\},
                                                    \label{eq:FACT}
\end{eqnarray}
in the BSW scheme. Factorizable amplitudes for the other 
$\bar B \rightarrow D\pi$ and $D^*\pi$ decays also can be calculated 
in the same way. To evaluate these amplitudes, we use the following 
parameterization of matrix elements of currents\cite{NRSX},  
\begin{equation}
       \langle \pi(q)|A_\mu^{}|0 \rangle = -if_{\pi}q_\mu, 
                                                   \label{eq:PCAC}
\end{equation}
\begin{eqnarray}
&&\langle D(p')|V_\mu^{}|\bar B(p)\rangle 
= \Biggl\{(p+p')_\mu - {m_B^2 - m_D^2 \over q^2}q_\mu\Biggr\}
F_1^{(D\bar B)}(q^2) 
        + {m_B^2 - m_D^2 \over q^2}q_\mu F_0^{(D\bar B)}(q^2),  
                                                    \label{eq:DB}
\\ &&\langle D^*(p')|A_\mu^{}|\bar B(p)\rangle 
= \Biggl\{
(m_B + m_{D^*})\epsilon_\mu^*(p')A_1^{(D^*\bar B)}(q^2) 
- {\epsilon^*(p')\cdot q \over m_B + m_{D^*}}(p + p')_\mu 
                                            A_2^{(D^*\bar B)}(q^2)
\nonumber \\
&& \hskip 50mm
- 2m_{D^*}{\epsilon^*\cdot q \over q^2}q_\mu A_3^{(D^*\bar B)}(q^2)
\Biggr\}
+ 2m_{D^*}{\epsilon^*\cdot q \over q^2}q_\mu A_0^{(D^*\bar B)}(q^2), 
                                                    \label{eq:D^*B}
\end{eqnarray}
where $q = p - p'$ and the form factors satisfy 
\begin{eqnarray}
&& A_3^{(D^*\bar B)}(q^2) 
= {m_B + m_{D^*} \over 2m_{D^*}}A_1^{(D^*\bar B)}(q^2) 
- {m_B - m_{D^*} \over 2m_{D^*}}A_2^{(D^*\bar B)}(q^2),        
                                                \label{eq:FFA}\\
&& 
F_1^{(D\bar B)}(0) = F_0^{(D\bar B)}(0), 
\qquad A_3^{(D^*\bar B)}(0) = A_0^{(D^*\bar B)}(0).   \label{eq:FF0}
\end{eqnarray}
To get rid of useless imaginary unit in the amplitude, however, we 
take the following parameterization of matrix element of vector 
current\cite{HY}, 
\begin{equation}
\langle V(p')|V_\mu^{}|0\rangle 
        = -if_{V}m_{V}\epsilon_{\mu}^*(p'),      \label{eq:Lvector}
\end{equation}
which can be treated in parallel to those of axial vector currents in 
Eq.(\ref{eq:PCAC}) in the IMF. Using these expressions of current 
matrix elements, we obtain the factorized amplitudes for the 
$\bar B \rightarrow D\pi$ and $D^*\pi$ decays in Table~1, where we 
have put $m_\pi^2=0$ and factored out the CKM matrix elements. 
As seen in Table~1, the color suppressed amplitudes are proportional 
to the small coefficient, $a_2$, and therefore they are suppressed. 

Before we evaluate numerically the factorized amplitudes, we study 
non-factorizale amplitudes for the $\bar B(p)\rightarrow D(p')\pi(q)$ 
[and $D^*(p')\pi(q)$] decays. The amplitudes for hadronic weak 
(two-body) decays with final state interactions will be described by 
diagrams similar to (b) and (c) in Fig.~1 in which soft gluon(s) are 
replaced by $\{q\bar q\}$ pair(s) [i.e., meson(s)], so that such 
amplitudes will be given by hadron dynamics and correspond to 
non-leading terms in the large $N_c$ expansion as mentioned before. 
Therefore, we assume that the non-factorizable amplitudes are 
dominated by dynamical contributions of various hadron states, and 
then calculate them using a hard pion technique with PCAC (partially 
conserved axial-vector 
\newpage
\begin{center}
\begin{quote}
{Table~1. Factorized amplitudes for $\bar B \rightarrow D\pi$ and 
$D^*\pi$ decays where $m_\pi^2=0$. The CKM matrix elements are
factored out.}
\end{quote}
\vspace{0.5cm}

\begin{tabular}
{l|l}
\hline\hline
\vspace{-4mm}\\
$\quad\,\,${\rm Decay}
&\hskip 5cm {$\quad A_{\rm FA}\,$}\\
 &\vspace{-4mm}\\
\hline 
 &\vspace{-4mm}\\
$\bar B^0 \rightarrow D^{+}\pi^-$
& $\displaystyle{
\,\quad i{G_F \over\sqrt{2}}a_1f_\pi(m_B^2 - m_D^2)F_0^{(D\bar B)}(0) 
\Biggl[
1 - \Biggl({a_2 \over a_1}\Biggr)\Biggl({f_B \over f_\pi}\Biggr)
\Biggl({m_D^2 \over m_B^2 - m_D^2}\Biggr)
{F_0^{(D\pi)}(m_B^2) \over F_0^{(D\bar B)}(0)}
\Biggr]}$
\\
 &\vspace{-4mm}\\
\hline
\vspace{-4mm}\\
$\bar B^0 \rightarrow D^{0}\pi^0$
& $\displaystyle{
\,-\,i{G_F \over {2}}a_2f_Dm_B^2F_0^{(\pi\bar B)}(m_D^2) 
\Biggl[
1 + \Biggl({f_B \over f_D}\Biggr)
\Biggl({m_D^2 \over m_B^2}\Biggr)
{F_0^{(D\pi)}(m_B^2) \over F_0^{(\pi\bar B)}(m_D^2)}
\Biggr]}$
\\
 &\vspace{-4mm}\\
\hline
\vspace{-4mm}\\
$B^- \rightarrow D^0\pi^-$
& $\displaystyle{\,\quad i
   {G_F \over\sqrt{2}}a_1f_\pi(m_B^2 - m_D^2)F_0^{(D\bar B)}(0) 
\Biggl[
1 + \Biggl({a_2 \over a_1}\Biggr)\Biggl({f_D \over f_\pi}\Biggr)
\Biggl({m_B^2 \over m_B^2 - m_D^2}\Biggr)
{F_0^{(\pi\bar B)}(m_D^2) \over F_0^{(D\bar B)}(0)}
\Biggr]}
$ 
\\
 &\vspace{-4mm}\\
\hline
\vspace{-4mm}\\
$\bar B^0 \rightarrow D^{*+}\pi^-$
& $\displaystyle{\,-\,i
   {G_F \over\sqrt{2}}a_1f_\pi A_0^{(D^*\bar B)}(0) 
\Biggl[
1 - \Biggl({a_2 \over a_1}\Biggr)\Biggl({f_B \over f_\pi}\Biggr)
{A_0^{(D^*\pi)}(m_B^2) \over A_0^{(D^*\bar B)}(0)}
\Biggr]2m_{D^*}\epsilon^*(p')\cdot p}$
\\
 &\vspace{-4mm}\\
\hline
\vspace{-4mm}\\
$\bar B^0 \rightarrow D^{*0}\pi^0$
& $\displaystyle{\,\quad i
   {G_F \over {2}}a_2f_{D^*}F_1^{(\pi\bar B)}(m_{D^*}^{2}) 
\Biggl[
1 + \Biggl({f_B \over f_{D^*}}\Biggr)
{A_0^{(D^*\pi)}(m_B^2) \over F_1^{(\pi\bar B)}(m_{D^*}^{2})}
\Biggr]2m_{D^*}\epsilon^*(p')\cdot p}$
\\
 &\vspace{-4mm}\\
\hline
\vspace{-4mm}\\
$B^- \rightarrow D^{*0}\pi^-$
& $\displaystyle{\,-\,i
   {G_F \over\sqrt{2}}a_1f_\pi A_0^{(D^*\bar B)}(0) 
\Biggl[
1 + \Biggl({a_2 \over a_1}\Biggr)\Biggl({f_{D^*} \over f_\pi}\Biggr)
{F_1^{(\pi\bar B)}(m_{D^*}^{2}) \over A_0^{(D^*\bar B)}(0)}
\Biggr]2m_{D^*}\epsilon^*(p')\cdot p}$ 
\vspace{-4mm}
\\
 & \\
\hline\hline
\end{tabular}
\end{center}
\vspace{10mm}
current) in the IMF; i.e., 
${\bf p} \rightarrow \infty$~\cite{hard pion,suppl},  
\begin{equation}
 M_{\rm NF}(\bar B\rightarrow D^{[*]}\pi) 
 \simeq \lim_{{\bf p} \to \infty,\,\, {\bf q} \to 0}
M_{\rm NF}(\bar B\rightarrow D^{[*]}\pi).
\end{equation}
In this approximation, $M_{\rm NF}$ is given by a sum of equal-time 
commutator (ETC) term and surface term, 
\begin{equation}
 M_{\rm NF}(\bar B \rightarrow D^{[*]}\pi)
\simeq M_{\rm ETC}(\bar B \rightarrow D^{[*]}\pi)
+ M_{\rm S}(\bar B \rightarrow D^{[*]}\pi).    \label{eq:hard pion}
\end{equation}
$M_{\rm ETC}$ has the same form as that in the old soft pion 
approximation~\cite{soft-pion}
\begin{equation}
M_{\rm ETC}(\bar B \rightarrow D^{[*]}\pi)
= {i \over f_{\pi}}
 \langle{D^{[*]}|[V_{\bar \pi}, \tilde H_w]|\bar B}\rangle.  
                                                     \label{eq:ETC}
\end{equation}
In the above equation, the commutation relation, 
$[V_\pi + A_\pi, \tilde H_w]=0$, has been used, where $V_\pi$ and 
$A_\pi$ are isospin charge and its axial counterpart, respectively. 
The surface term, $M_S$, is given by 
\begin{equation}
M_{\rm S}(\bar B \rightarrow D^{[*]}\pi)
=  \lim_{{\bf p} \to \infty,\,\,{\bf q} \to 0}
\Biggl\{-{i \over f_\pi}q^\mu T_\mu^{[*]}\Biggr\}, 
\end{equation}
with the hypothetical amplitude 
\begin{equation}
T_\mu^{[*]} = i\int e^{iqx}
{\langle{D^{[*]}(p')|T[A_\mu^{(\bar\pi)}\tilde H_w]|\bar B(p)}\rangle}
                                                                 d^4x.
\end{equation}
When a complete set of energy eigen states is inserted between 
products of axial vector current and $\tilde H_w$ in $T_\mu^{[*]}$, 
contributions of single meson intermediate states survive in contrast 
with the soft pion approximation. In this way, $M_{\rm S}$ is 
provided by a sum of all possible pole amplitudes, 
\begin{equation}
M_{\rm S} = \sum_n M_{\rm S}^{(n)} + \sum_l M_{\rm S}^{(l)}. 
                                               \label{eq:surf-tot}
\end{equation}
$M_{\rm S}^{(n)}$ and $M_{\rm S}^{(l)}$ are pole amplitudes in 
the $s$- and $u$-channels, respectively, and given by  
\begin{eqnarray} 
&&M_{\rm S}^{(n)}(\bar B \rightarrow D^{[*]}\pi) 
= -{i \over f_{\pi}}\Bigl({m_{D^{[*]}}^2 - m_B^2 
                                         \over m_n^2 - m_B^2}\Bigr)
  \langle{D^{[*]}|A_{\bar \pi}|n}\rangle
               \langle{n|\tilde H_w|\bar B}\rangle,  \nonumber \\
&& M_{\rm S}^{(l)}\,(\bar B \rightarrow D^{[*]}\pi) 
= -{i \over f_{\pi}}\Bigl({m_{D^{[*]}}^2 - m_B^2 
                              \over m_\ell^2 - m_{D^{[*]}}^2}\Bigr)
\langle{D^{[*]}|\tilde H_w|l}\rangle
                  \langle{l|A_{\bar \pi}|\bar B}\rangle.  
                                                    \label{eq:SURF}
\end{eqnarray}
Therefore, $M_{\rm S}$ provides corrections to the soft pion 
approximation, i.e., the present hard pion technique can be considered 
as an innovation of the old soft pion technique. (See, for more
details and notations, Refs.\cite{hard pion} and \cite{suppl}.) $n$ 
and $l$ in Eq.(\ref{eq:SURF}) run over all possible single meson 
states, not only ordinary $\{q\bar q\}$, but also hybrid 
$\{q\bar qg\}$, four-quark $\{qq\bar q\bar q\}$, glue-balls, etc. 
However, $|n\rangle$ and $|l\rangle$ as well as the external states
are energy eigen states in the present case. Since the states which 
sandwich $\tilde H_w$ should conserve their spins in the rest frame
and since Lorentz invariant amplitudes are considered, only the states 
$|n\rangle$ and $|l\rangle$ which conserve spins in the matrix
elements, 
$\langle{n|\tilde H_w|\bar B}\rangle$ and 
$\langle{D^{[*]}|\tilde H_w|l}\rangle$, 
should be picked up~\cite{Pakvasa}. Therefore, we discard vector 
meson pole contributions to the $u$-channel of 
pseudo-scalar(PS)-meson decays into two PS-meson states although we
have taken such contributions in our previous papers~\cite{VMP}. 
Since the $B$ meson mass $m_B$ is much higher than those of charm 
mesons and since wave function overlappings between the ground-state 
$\{q\bar q\}_0$ and excited-state-meson states are expected to be 
small, however, excited meson contributions will be small in these 
decays and can be safely neglected. Therefore the hard pion 
amplitudes as the non-factorizable long distance ones are 
approximately described in terms of 
{\it asymptotic ground-state-meson matrix elements} (matrix elements 
taken between single ground-state-meson states with infinite 
momentum) of $V_\pi$, $A_\pi$ and $\tilde H_w$. 

Amplitudes for dynamical hadronic processes are usually decomposed into 
({\it continuum contribution}) + ({\it  Born term}).  
Since $M_{\rm S}$ has been given by a sum of pole amplitudes, 
$M_{\rm ETC}$ corresponds to the continuum contribution\cite{MATHUR} 
which can develop a phase relative to the Born term. Therefore, we 
parameterize the ETC terms using isospin eigen amplitudes and their 
phases. Since the $D\pi$ final state can have isospin 
$I={1 \over 2}$ and ${3 \over 2}$, 
$M_{\rm ETC}(\bar B\rightarrow D\pi)$'s are parameterized as  
\begin{eqnarray}
&&M_{\rm ETC}(\bar B^0\, \rightarrow D^+\pi^-) 
= \,\,\,\,\sqrt{{1 \over 3}}M_{\rm ETC}^{(3)}e^{i\tilde\delta_{3}} 
+ \sqrt{{2 \over 3}}M_{\rm ETC}^{(1)}e^{i\tilde\delta_{1}},  
                                               \label{eq:ETCMP}\\
&&M_{\rm ETC}(\bar B^0\, \rightarrow D^0\,\pi^0\,) 
= -\sqrt{{2 \over 3}}M_{\rm ETC}^{(3)}e^{i\tilde\delta_{3}} 
+  \sqrt{{1 \over 3}}M_{\rm ETC}^{(1)}e^{i\tilde\delta_{1}},  
                                                \label{eq:ETCZZ}\\
&&M_{\rm ETC}(B^- \rightarrow  D^0\,\pi^-) 
= \,\,\,\,\,\sqrt{3}M_{\rm ETC}^{(3)}e^{i\tilde\delta_{3}},            
                                                \label{eq:ETCZP}
\end{eqnarray}
where $M_{\rm ETC}^{(2I)}$'s are the isospin eigen amplitudes with
isospin $I$ and $\tilde\delta_{2I}$'s are the corresponding phase
shifts introduced. In the present approach, therefore, the final
state interactions are included in the non-factorizable amplitudes. 
This is compatible with the fact that amplitudes with final state 
interactions are given by diagrams which belong to non-leading terms 
in the large $N_c$ expansion. 

Asymptotic matrix elements of $V_{\pi}$ and $A_{\pi}$ are
parameterized as 
\begin{eqnarray}
&&\langle{\pi^0|V_{\pi^+}|\pi^-}\rangle 
    = \sqrt{2}\langle{K^{+}|V_{\pi^+}|K^0}\rangle 
    = -\sqrt{2}\langle{D^{+}|V_{\pi^+}|D^0}\rangle 
    =  \sqrt{2}\langle{B^{+}|V_{\pi^+}|B^0}\rangle 
    = \cdots = \sqrt{2},                          \label{eq:MEV}\\
&&\langle{\rho^0|A_{\pi^+}|\pi^-}\rangle 
    = \sqrt{2}\langle{K^{*+}|A_{\pi^+}|K^0}\rangle 
    = -\sqrt{2}\langle{D^{*+}|A_{\pi^+}|D^0}\rangle 
    =  \sqrt{2}\langle{B^{*+}|A_{\pi^+}|B^0}\rangle 
    = \cdots = h.                                     \label{eq:MEA}
\end{eqnarray}
The above parameterization can be obtained by using asymptotic 
$SU_f(5)$ symmetry\cite{ASYMP}, or $SU_f(5)$ extension of the nonet 
symmetry in $SU_f(3)$. Asymptotic matrix elements of $V_\pi$ between 
vector meson states can be obtained by exchanging PS-mesons for 
vector mesons with corresponding flavors in Eq.(\ref{eq:MEV}), for 
example, as $\pi^{0,-} \rightarrow \rho^{0,-}$, etc. The $SU_f(4)$ 
part of the above parameterization reproduces 
well\cite{suppl,Takasugi} the observed values of decay rates, 
$\Gamma(D^*\rightarrow D\pi)$ and $\Gamma(D^*\rightarrow D\gamma)$. 

In this way, we can describe the non-factorizable amplitudes for the 
$\bar B \rightarrow D\pi$ decays as 
\begin{eqnarray}
&&M_{\rm NF}(\bar B^0\, \rightarrow D^+\pi^-) 
\simeq -i{\langle{D^0|\tilde H_w|\bar B^0}\rangle \over f_\pi}
\Biggl\{
\Biggl[{4\over 3}e^{i\tilde\delta_1} 
                        -{1\over 3}e^{i\tilde\delta_3}\Biggr]
                            +  \cdots  \Biggr\},  \label{eq:LMP}
\\
&&M_{\rm NF}(\bar B^0\, \rightarrow D^0\,\pi^0\,) 
\simeq -i{\langle{D^0|\tilde H_w|\bar B^0}\rangle \over f_\pi}
\Biggl\{
{\sqrt{2} \over 3}\Biggl[2e^{i\tilde\delta_1} 
                                  + e^{i\tilde\delta_3}\Biggr]
                               + \cdots  \Biggr\},  \label{eq:LZZ}
\\
&&M_{\rm NF}(B^- \rightarrow D^0\,\pi^-) 
\simeq\,\,\,\, 
       i{\langle{D^{0}|\tilde H_w|\bar B^0}\rangle \over f_\pi}
                                       \Biggl\{e^{i\tilde\delta_3} 
                                 +\cdots\Biggr\},   \label{eq:LZP}
\end{eqnarray}
where the ellipses denote the neglected pole contributions. 

In the case of the $\bar B \rightarrow D^*\pi$ decays, the matrix 
element ${\langle V|\tilde H_w|P\rangle}$ should vanish because of 
conservation of spin as discussed before, so that 
$M_{\rm ETC}(\bar B \rightarrow D^*\pi)$ also should vanish but now 
$D$ and $B^*$ poles in the $s$- and $u$-channels, respectively, 
survive, i.e., 
\begin{eqnarray}
&&M_{\rm NF}(\bar B^0\, \rightarrow D^{*+}\pi^-) 
\simeq {i\over f_\pi}
\langle{D^{0}|\tilde H_w|\bar B^{0}}\rangle 
\Biggl({m_{B}^2-m_{D^*}^2 \over m_{B}^2-m_{D}^2}\Biggr)               
          \sqrt{1\over 2}h + \cdots ,              \label{eq:LSMP}
\\
&&M_{\rm NF}(\bar B^0\, \rightarrow D^{*0}\,\pi^0\,) 
\simeq {i \over \sqrt{2}f_\pi}
\Biggl[\langle{D^{0}|\tilde H_w|\bar B^0}\rangle 
         \Biggl({m_{B}^2-m_{D^*}^2 \over m_{B}^2-m_{D}^2}\Biggr)
\nonumber \\
&&\hspace{60mm}+ \langle{D^{*0}|\tilde H_w|\bar B^{*0}}\rangle 
\Biggl({m_{B}^2-m_{D^*}^2 \over m_{B^*}^2-m_{D^*}^2}\Biggr)\Biggr]
              \sqrt{1\over 2}h +\cdots,  \label{eq:LSZZ}\\
&&M_{\rm NF}(B^- \rightarrow D^{*0}\,\pi^-) 
\simeq -{i \over f_\pi}\langle{D^{*0}|\tilde H_w|\bar B^{*0}}\rangle
\Biggl({m_{B}^2-m_{D^*}^2 \over m_{B^*}^2-m_{D^*}^2}\Biggr)
           \sqrt{1\over 2}h +\cdots,   \label{eq:LSZP}
\end{eqnarray}
where the ellipses denote the neglected excited meson contributions. 
Therefore the non-factorizable amplitudes in the hard pion 
approximation are controlled by the asymptotic ground-state-meson 
matrix elements of $\tilde H_w$ (and the possible phases). 

Now we evaluate the above amplitudes in the IMF. The factorized
amplitudes in Table~1 contain many parameters which have not been 
measured by experiments, {\it i.e.}, form factors, 
$F^{(D\bar B)}_0(q^2)$, $A^{(D^*\bar B)}_0(q^2)$, 
$F^{(\pi\bar B)}_0(q^2)$, $F^{(\pi\bar B)}_1(q^2)$, etc., 
and decay constants, $f_D$, $f_D^*$, $f_B$, etc.  The form factors 
$F_0^{(D\bar B)}(0)$ and $A_0^{(D^*\bar B)}(0)$ have been calculated 
by using the heavy quark effective theory (HQET)~\cite{HQET}. 
The other form factors are concerned with light mesons and therefore 
have to be estimated by some other models. In color favored decays, 
however, main parts of the factorized amplitudes depend on the form 
factor, $F_0^{(D\bar B)}(0)$ or $A_0^{(D^*\bar B)}(0)$, and the other 
form factors are included in minor terms proportional to $a_2$. 
Therefore our result may not be lead to serious uncertainties 
although some model dependent values of the form factors are taken. 
(We will take, later, 
$F_0^{(D\bar B)}(0)\simeq A_0^{(D^*\bar B)}(0)\simeq 0.59$ 
as in Ref.\cite{Kamal}, 
$F^{(\pi\bar B)}_0(m_D^2)\simeq 0.30$ and 
$F^{(\pi\bar B)}_1(m_{D^*}^2)\simeq 0.34$ 
as expected in pQCD~\cite{pQCD}.) In the color suppressed 
$\bar B^0\rightarrow D^0\pi^0$ and $D^{*0}\pi^0$ decays, however, 
the factorized amplitudes are proportional to the form factors, 
$F_0^{(\pi\bar B)}(m_D^2)$ and $F_1^{(\pi\bar B)}(m_{D^*}^{2})$, 
respectively. Since their values are model dependent, the result on 
the color suppressed decays may be a little ambiguous, if 
non-factorizable contribution is less important. For the decay 
constants of heavy mesons, we assume $f_D \simeq f_{D^*}$ (and 
$f_B \simeq f_{B^*}$) since $D$ and $D^*$ ($B$ and $B^*$) are 
expected to be degenerate because of heavy quark symmetry~\cite{HQET} 
and are approximately degenerate in reality. Here we take 
$f_{D^*} \simeq f_D \simeq 211$ MeV and 
$f_{B^*} \simeq f_B \simeq 179$ MeV 
from a recent result of lattice QCD\cite{Lattice}. In this way, we 
can obtain the factorized amplitudes in the second column of Table~2, 
where we have neglected very small annihilation terms in the 
$\bar B^0 \rightarrow D^0\pi^0$ and $D^{*0}\pi^0$ decay amplitudes. 

To evaluate the non-factorizable amplitudes, we need to know the size 
of the asymptotic matrix elements of $\tilde H_w$ and $A_\pi$. The 
asymptotic matrix elements of $A_\pi$ which was parameterized in 
Eq.(\ref{eq:MEA}) is estimated~\cite{hard pion,suppl} to be 
$|h|\simeq 1.0$ by using PCAC and the observed rate\cite{PDG00}, 
$\Gamma(\rho \rightarrow \pi\pi)_{\rm exp} \simeq 150$ MeV. 
We here take its positive sign, i.e., $h\simeq 1.0$, since, in the 
IMF, $\langle{\rho^0|A_{\pi^+}|\pi^-}\rangle$ is given by the form 
factor $A_3^{(\rho\pi)}(0)$ included in the matrix element of 
axial-vector current. In this way, the factorized and 
non-factorizable amplitudes for the $B^-\rightarrow D^{*0}\pi^-$ 
decay interfere constructively with each other. For the asymptotic 
matrix elements, 
$\langle D^0|\tilde H_w|\bar B^0 \rangle$ and 
$\langle D^{*0}|\tilde H_w|\bar B^{*0} \rangle$, 
we treat them as unknown parameters and search phenomenologically for
their values to reproduce the observed rates for the 
$\bar B\rightarrow D^{[*]}\pi$ decays. To this, we assume 
\begin{equation}
\langle {D^{*0}|\tilde H_w|\bar B^{*0}} \rangle 
\simeq \langle {D^{0}|\tilde H_w|\bar B^{0}} \rangle
                                           \label{eq:AME-HQS}
\end{equation}
as expected in the heavy quark symmetry and parameterize these 
matrix elements using factorizable ones of $H_w^{\rm BSW}$ as 
\begin{equation}
\langle {D^{0}|\tilde H_w|\bar B^{0}} \rangle 
=B_H \langle {D^{0}|H_w^{\rm BSW}|\bar B^{0}} \rangle_{\rm FA}
                                           \label{eq:AME-Hw}
\end{equation}
with 
\begin{equation}
\langle {D^{0}|H_w^{\rm BSW}|\bar B^{0}} \rangle _{FA} 
={G_F\over \sqrt{2}}V_{cb}V_{ud}
\Bigl({m_D^2 + m_B^2 \over 2}\Bigr) f_Df_Ba_2,
                                            \label{eq:AME-Hw-FA}
\end{equation}
where $B_H$ is a parameter introduced. In this way, we obtain the 
hard pion amplitudes as the non-factorizable contributions listed 
in the third column of Table~2, where the CKM matrix elements have 
been factored out. 

Before we compare our result with the observations, we study
phenomenologically allowed branching ratios for the 
$\bar B\rightarrow D\pi$ and $D^{*}\pi$ decays. For the color
suppressed $\bar B^0 \rightarrow D^0\pi^0$ and $D^{*0}\pi^0$ 
decays, in particular, only the upper limits of their branching 
ratios have been 
\newpage
\begin{center}
\begin{quote}
{Table~2. Factorized and non-factorizable amplitudes for the 
$\bar B \rightarrow D\pi$ and $D^*\pi$ decays.
The CKM matrix elements are factored out. }
\end{quote}
\vspace{0.5cm}

\begin{tabular}
{l|l|l}
\hline\hline
\vspace{-4mm}\\
$\quad\,\,${\rm Decay}
&$\quad A_{\rm FA}\,(\times 10^{-5}$  GeV)
&$\qquad\qquad A_{\rm NF}\,(\times 10^{-5}$ GeV) 
\\
 & &\vspace{-4mm}\\
\hline 
 & &\vspace{-4mm}\\
$\bar B^0 \rightarrow D^{+}\pi^-$
& $\displaystyle{\quad 1.57\,a_1\Bigl\{
                  1 - 0.03\Bigl({a_2 \over a_1}\Bigr)\Bigr\}}$
&$\displaystyle{\, -3.70a_2B_H\,\Bigl\{
\Bigl[{4\over 3}e^{i\tilde\delta_1} 
                   -{1\over 3}e^{i\tilde\delta_3}\Bigr] 
                                              \Bigr\}}$
\\
 & &\vspace{-4mm}\\
\hline 
 & &\vspace{-4mm}\\
$\bar B^0 \rightarrow D^{0}\pi^0$
& $\displaystyle{\, -1.03\,a_2\Bigl\{
                         {f_D \over 0.211\,\,{\rm GeV}}\Bigr\}}$
& $\displaystyle{\, -3.70a_2B_H\,\Bigl\{
{\sqrt{2} \over 3}\Bigl[2e^{i\tilde\delta_1} 
                                   + e^{i\tilde\delta_3}\Bigr]
                                              \Bigr\}}$
\\
 & &\vspace{-4mm}\\
\hline 
 & &\vspace{-4mm}\\
$B^- \rightarrow D^0\pi^-$
& $\displaystyle{\quad 1.57\,a_1\Bigl\{
                   1 + 0.94\Bigl({a_2 \over a_1}\Bigr)\Bigr\}}$ 
& $\displaystyle{\quad 3.70a_2B_H
                          \,\Bigl\{e^{i\tilde\delta_3}\Bigr\}}$
\\
 & &\vspace{-4mm}\\
\hline 
 & &\vspace{-4mm}\\
$\bar B^0 \rightarrow D^{*+}\pi^-$
& $\displaystyle{\, -1.53\,a_1\Bigl\{
                    1 - 0.22\Bigl({a_2 \over a_1}\Bigr)\Bigr\}}$
& $\displaystyle{\, - 3.70a_2B_H\,\Bigl\{ 
  0.694 \Bigr\}}$
\\
 & &\vspace{-4mm}\\
\hline 
 & &\vspace{-4mm}\\
$\bar B^0 \rightarrow D^{*0}\pi^0$
& $\displaystyle{\quad 0.997\,a_2\Bigl\{{f_{D^*} 
                           \over 0.211\,\,{\rm GeV}}\Bigr\}}$
& $\displaystyle{\quad 3.70a_2B_H\,\Bigl\{
  0.983 \Bigr\}}$
\\
 & &\vspace{-4mm}\\
\hline 
 & &\vspace{-4mm}\\
$B^- \rightarrow D^{*0}\pi^-$
& $\displaystyle{\, -1.53\,a_1\Bigl\{
                 1 + 0.92\Bigl({a_2 \over a_1}\Bigr)\Bigr\}}$ 
& $\displaystyle{\,- 3.70a_2B_H\,\Bigl\{
  0.696 \Bigr\}}$
\vspace{-4mm}
\\
 & & \\
\hline\hline
\end{tabular}

\end{center}
\vspace{10mm}
given as their measured values at the present stage. 
However, we can estimate their lower bounds compatible with the 
measured branching ratios for the charged modes. To this, we 
parameterize the amplitudes for these decays as 
\begin{eqnarray}
&&A(\bar B^0\, \rightarrow D^{[*]+}\pi^-) 
= \,\,\,\,\sqrt{{1 \over 3}}A^{[*]}_{3}e^{i\delta_{3}^{[*]}} 
+ \sqrt{{2 \over 3}}A^{[*]}_{1}e^{i\delta_{1}^{[*]}},  
                                               \label{eq:amp-pm}\\
&&A(\bar B^0\, \rightarrow D^{[*]0}\,\pi^0\,) 
= -\sqrt{{2 \over 3}}A^{[*]}_{3}e^{i\delta_{3}^{[*]}} 
+  \sqrt{{1 \over 3}}A^{[*]}_{1}e^{i\delta_{1}^{[*]}},  
                                               \label{eq:amp-zz}\\
&&A(B^- \rightarrow  D^{[*]0}\pi^-) 
= \quad\sqrt{3}A^{[*]}_{3}e^{i\delta_{3}^{[*]}},            
                                                \label{eq:amp-zm}
\end{eqnarray}
since the $D\pi$ [and $D^*\pi$] final states can have $I={1\over 2}$ 
and ${3\over 2}$, where $A^{[*]}_{2I}$'s and $\delta_{2I}^{[*]}$'s 
are isospin eigen amplitudes for the $\bar B\rightarrow D^{[*]}\pi$ 
decays and their phases, respectively. Taking positive values of the 
ratio of isospin eigen amplitudes, $r^{[*]}=A_{3}^{[*]}/A^{[*]}_{1}$, 
we obtain 
\begin{equation}
\cos(\delta^{[*]}) 
= \Bigl({9R_{0-}^{[*]} - 1\over 4}\Bigr)r^{[*]} - {1\over r^{[*]}}, 
                                           \label{eq:phase-diff}
\end{equation}
where 
\begin{equation}
\delta^{[*]} = \delta_1^{[*]} -\delta_3^{[*]} \quad {\rm and} \quad 
r^{[*]} = \sqrt{1\over
                      3R_{00}^{[*]}R_{0-}^{[*]}+3R_{0-}^{[*]}-1}\,\,.     
                                              \label{eq:ratio-amps}
\end{equation} 
Here $R_{0-}^{[*]}$ and $R_{00}^{[*]}$ are ratios of decay rates, 
\begin{equation}
R_{0-}^{[*]} = {\Gamma(\bar B^0\rightarrow D^{[*]+}\pi^-) 
\over\Gamma(B^-\rightarrow D^{[*]0}\pi^-)} 
\quad {\rm and}\quad
R_{00}^{[*]} = {\Gamma(\bar B^0\rightarrow D^{[*]0}\pi^0) 
\over \Gamma(\bar B^0\rightarrow D^{[*]+}\pi^-)}.
                                           \label{eq:ratio-rates}
\end{equation}
\newpage
\begin{quote}
{Table~3. Branching ratios ($\%$) for $\bar B \rightarrow D\pi$ 
and $D^*\pi$ decays, where the values of the CKM matrix elements, 
$V_{cb}=0.040$ and $V_{ud}=0.97$, the lifetimes, 
$\tau(B^-) = 1.65\times10^{-12}$~s and 
$\tau(\bar B^0) = 1.55\times10^{-12}$~s 
from the updated experimental data~\cite{PDG00}, $a_1=1.024$ with the 
LO QCD corrections and phenomenological $a_2=0.21$, $B_H=0.17$, 
$\tilde\delta_1=88^\circ$ and $\tilde\delta_3=8^\circ$ are taken. 
${\cal B}_{\rm FA}$ and ${\cal B}_{\rm tot}$ are given by 
the factorized amplitude and a sum of the factorized and 
non-factorizable ones, respectively. The values of 
${\cal B}_{\rm ph}$ are estimated in the text. 
} 
\end{quote}

\begin{center}
\begin{tabular}
{|c|l|l|c|c|}
\hline\hline
 & & & & \vspace{-4mm}\\
 {Decays} 
&\quad ${\cal B}_{\rm FA}$ \quad
&\quad ${\cal B}_{\rm tot}$ \quad
&\quad ${\cal B}_{\rm ph}$ \quad
&\quad ${\cal B}_{\rm exp}$ ($\ast$) \qquad
\\
 & & & &\vspace{-4mm}\\
\hline
\vspace{-4mm}\\
{$\bar B^0\rightarrow D^+\pi^-$}
&\hspace{2mm}{0.29}
&\hspace{2mm}{0.30}
&$0.28 - 0.34$
&$0.30 \pm 0.04$
\\
 & & & &\vspace{-4mm}\\
\hline
\vspace{-4mm}\\
{$\bar B^0\rightarrow D^0\pi^0$}
&\hspace{2mm}{0.005}
&\hspace{2mm}{0.011}
&\hspace{2mm}$0.006 - 0.012$\hspace{2mm}
&$< 0.012$
\\
 & & & &\vspace{-4mm}\\
\hline 
\vspace{-4mm}\\
{$B^-\rightarrow D^0\pi^-$}
&\hspace{2mm}{0.45}
&\hspace{2mm}{0.52}
&$0.53 \pm 0.05$
&$0.53 \pm 0.05$
\\
 & & & &\vspace{-4mm}\\
\hline
\vspace{-4mm}\\
{$\bar B^0\rightarrow D^{*+}\pi^-$}
&\hspace{2mm}{0.25}
&\hspace{2mm}{0.28}
&\hspace{2mm}$0.276 \pm 0.021$
&\hspace{2mm}$0.276 \pm 0.021$
\\
 & & & &\vspace{-4mm}\\
\hline
\vspace{-4mm}\\
{$\bar B^0\rightarrow D^{*0}\pi^0$}
&\hspace{2mm}{0.005}
&\hspace{2mm}{0.013}
&\hspace{2mm}$0.004 - 0.044$
&$< 0.044$
\\
 & & & &\vspace{-4mm}\\
\hline
\vspace{-4mm}\\
{$B^-\rightarrow D^{*0}\pi^-$}
&\hspace{2mm}{0.42}
&\hspace{2mm}{0.46}
&\hspace{2mm}$0.46 \pm 0.04$
&\hspace{2mm}$0.46 \pm 0.04$
\vspace{-4mm}\\
 & & & &\\
\hline\hline
\end{tabular}\vspace{3mm}\\
($\ast$) The data values are taken from Ref.~\cite{PDG00}. 
\end{center}
\vspace{5mm}
Values of $R_{0-}^{[*]}$ and $R_{00}^{[*]}$ can be estimated 
phenomenologically from the experimental data~\cite{PDG00} on  
branching ratios for $\bar B\rightarrow D^{[*]}\pi$ decays in 
Table~3 as 
$R_{0-} = 0.58 \pm 0.10,\,\, R_{00} < 0.04$ and 
$R_{0-}^{*} = 0.61 \pm 0.08,\,\, R_{00}^{*} < 0.16$. 
However, these values of $R_{0-}$ and $R_{00}$ [$R_{0-}^{*}$ and 
$R_{00}^{*}$] are not always compatible with each other, i.e., 
the right-hand-side (r.h.s.) in Eq.(\ref{eq:phase-diff}) is not 
always less than unity. It is satisfied in more restricted regions 
of $R$ and $R^{*}$, i.e., approximately, 
$0.04 \gtrsim R_{00} \gtrsim 0.02$ and 
$0.68 \gtrsim  R_{0-} \gtrsim 0.61$  
for the $\bar B\rightarrow D\pi$ decays, and 
$0.16 \gtrsim R^*_{00} \gtrsim 0.02$ and  
$0.69 \gtrsim R^*_{0-}\gtrsim 0.53$ 
for the $\bar B\rightarrow D^*\pi$ decays. These values of $R$ and 
$R^{*}$ lead approximately to the phenomenological branching ratios, 
${\cal B}_{\rm ph}$'s in Table~3, when the experimental data, 
${\cal B}(B^- \rightarrow D^0\pi^-)_{\rm exp}= 0.53 \pm 0.05$ and 
${\cal B}(B^- \rightarrow D^{*0}\pi^-)_{\rm exp}= 0.46 \pm 0.04$, 
are fixed. Here we put 
${\cal B}(\bar B^0 \rightarrow D^{*+}\pi^-)_{\rm ph} 
={\cal B}(\bar B^0 \rightarrow D^{*+}\pi^-)_{\rm exp}$ 
since $\cos(\delta^{*}) \leq 1$ is satisfied for all the 
experimentally allowed values of $R_{0-}^*$. The allowed values of 
$\cos(\delta)$ are limited within a narrow region, 
$0.96 \lesssim \,\cos(\delta)\, \leq \,1$, 
in which $\cos(\delta)$ is very close to unity while $\cos(\delta^*)$ 
is a little more mildly restricted (at the present stage) compared 
with the above $\cos(\delta)$, i.e., approximately, 
$0.70 \lesssim \cos(\delta^{*}) \leq\,\, 1 $. 
Therefore, the phase difference between the $I={1\over 2}$ and 
${3\over 2}$ amplitudes for the $\bar B \rightarrow D\pi$ decays is 
very small while, in the $\bar B \rightarrow D^*\pi$ decays, the 
allowed region of the corresponding phase difference is broader at 
the present stage. 

We now compare our result on the branching ratios, 
${\cal B}(\bar B\rightarrow D\pi)$ and 
${\cal B}(\bar B \rightarrow D^*\pi)$, 
with their phenomenologically estimated values (and experimental 
data), taking a sum of the factorized amplitude and the 
non-factorizable amplitude as the total one. To this, we determine 
values of parameters involved. We take $V_{cb}=0.040$ from the 
updated value $|V_{cb}|=0.0402 \pm 0.0019$\cite{PDG00}. For the 
coefficients $a_1$ and $a_2$ in $H_w^{\rm BSW}$, we do not know their 
true values. It has been known~\cite{BURAS} that the NLO corrections 
to $a_1$ are small while the corresponding ones to $a_2$ may be not 
much smaller compared with the LO corrections and depend strongly on 
the renormalization scheme. Therefore, we expect that the value, 
$a_1=1.024$ with the LO corrections at the scale 
$\mu\simeq m_b$~\cite{BURAS}, is not very far from the true value, 
and hence we take the above value of $a_1$. For $a_2$, however, we 
consider two cases. As the first case (i), we take $a_2=0.125$ with 
the LO QCD corrections at $\mu\simeq m_b$~\cite{BURAS}, and then we 
treat it as an adjustable parameter around the above value of $a_2$ 
as the second case (ii). For the phases $\tilde\delta_1$ and 
$\tilde\delta_3$ arising from contributions of non-resonant 
multi-hadron intermediate states into isospin $I={1\over 2}$ and 
${3\over 2}$ final states, they are restricted in the region 
$|\tilde\delta_{2I}| < 90^\circ$ since resonant contributions have 
already been extracted as pole amplitudes in $M_{\rm S}$ although 
their contributions are neglected as discussed before. For $B_H$, 
we here treat it as a free parameter but expect to be less than
unity. 

We now search for values of parameters, $\tilde\delta_1$,
$\tilde\delta_3$ and $B_H$ in the case (i), and $a_2$, 
$\tilde\delta_1$, $\tilde\delta_3$ and $B_H$ in the case (ii), which 
reproduce the phenomenologically estimated branching ratios (from the 
observed ones) for the $\bar B\rightarrow D^{[*]}\pi$ decays. In the
case (i), it is hard to reproduce them simultaneously. In the case
(ii), however, large $\tilde\delta_1$, 
($90^\circ > \tilde\delta_1\gtrsim 60^\circ$), and small 
$|\tilde\delta_3|$ are favored (but our result is not very 
sensitive to the latter). For $a_2$ and $B_H$, larger values of 
$a_2$, ($0.26\gtrsim a_2 \gtrsim 0.16$), compared with $a_2=0.125$ 
with the LO QCD corrections and rather small values of $B_H$, 
($0.07 \lesssim B_H \lesssim 0.28$), are favored. We list our results 
on the branching ratios for 
$a_1=1.024$, $a_2=0.21$, $\tilde\delta_1=88^\circ$, 
$\tilde\delta_3=8^\circ$ and $B_H=0.17$ 
in Table~3, where we have used the values, 
$V_{cb}=0.040$, $V_{ud}=0.97$, 
$\tau(B^-) = 1.65\times10^{-12}$~s and 
$\tau(\bar B^0) = 1.55\times10^{-12}$~s 
from the updated experimental data~\cite{PDG00}. ${\cal B}_{\rm FA}$ 
and ${\cal B}_{\rm tot}$ are given by the factorized amplitude and a 
sum of the factorized and non-factorizable ones, respectively. Values 
of ${\cal B}_{\rm ph}$ have been obtained phenomenologically from 
${\cal B}_{\rm exp}$ before. ${\cal B}_{\rm FA}$, in which the 
non-factorizable contributions are discarded, reproduces fairly well 
the existing data. However, if we add the non-factorizable 
contributions, we can improve the fit to the phenomenologically 
estimated ${\cal B}_{\rm ph}$. It is seen that the non-factorizable 
contributions to the color favored 
$\bar B \rightarrow D\pi$ and $D^*\pi$ decays are rather small but 
still can interfere efficiently with the main amplitude given by the 
naive factorization. 

\section{$\bar B \rightarrow J/\psi\bar K$ and $J/\psi\pi$ decays}

Now we study CKM-angle favored $\bar B \rightarrow J/\psi\bar K$ 
and suppressed $\bar B\rightarrow J/\psi\pi$ decays in the same way 
as in the previous section. Both of them are color mismatched and 
their kinematical condition is much different from the color favored 
$\bar B \rightarrow D\pi$ and $D^*\pi$ decays at the level of 
underlying quarks, i.e., 
$b \rightarrow (c\bar c)_1\,+\,s\, ({\rm or}\,\,d)$ 
in the former but $b \rightarrow c\,+\,(\bar ud)_1$ in the latter. 
Therefore, if the large $N_c$ argument does not work well in hadronic
weak decays as seen before, dominance of factorized amplitudes in
these decays cannot be guaranteed and hence non-factorizable long
distance contribution can play an important role. 

The factorized amplitude for the 
$\bar B \rightarrow J/\psi\bar K$ decays is given by 
\begin{equation}
M_{\rm FA}(\bar B \rightarrow J/\psi\bar K)
=-iV_{cb}V_{cs}
\Bigl\{
{G_F \over \sqrt{2}}a_2f_\psi F_1^{(\bar K\bar B)}(m_\psi^2)
\Bigr\}2m_\psi\epsilon^*(p')\cdot p. 
\end{equation}
The value of the decay constant of $J/\psi$ is estimated to be 
$f_\psi \simeq 406$~MeV from the observed 
rate\cite{PDG00} for the 
$J/\psi \rightarrow \ell^+\ell^-$. The value of the CKM matrix 
element $V_{cs}$ is given by 
\newpage 
\begin{quote}
{Table~4. Branching ratios ($\%$) for the 
$\bar B \rightarrow J/\psi\bar K$ decays,  
where the values of $F_1^{(\bar K\bar B)}(m_\psi^2)$ from the models, 
BSW, CDDFGN and GKP, in Refs.~\cite{BSW}, \cite{CDDFGN} and \cite{GKP},
respectively, are used. Values of the other parameters involved are
the same as in Table~3, where  $B_H'=B_H$ is assumed. The data 
values are taken from Ref.~\cite{PDG00}.}
\end{quote}

\begin{center}
\begin{tabular}
{|c|c|c|c|}
\hline\hline
 & & & \vspace{-4mm}\\
{$\hspace{3mm}$Models}
& $\qquad$BSW$\qquad$
& $\quad$CDDFGN$\quad$
& $\quad$GKP$\quad$
\\
 & & & \vspace{-4mm}\\
\hline
\vspace{-4mm}\\
{$\quad F_1^{(\bar K\bar B)}(m_\psi^2)\quad$}
&$\quad$0.565$\quad$
&$\quad$0.726$\quad$
&$\quad$0.837$\quad$
\\
 & & & \vspace{-4mm}\\
\hline
\vspace{-4mm}\\
{$\quad {\cal B}_{\rm FA}(\bar B\rightarrow J/\psi\bar K)
                                                \,(\%) \quad$}
&{$\quad$0.048$\quad$}
&{$\quad$0.079$\quad$}
&{$\quad$0.10$\quad$}
\\
 & & & \vspace{-4mm}\\
\hline
\vspace{-4mm}\\
$\quad {\cal B}_{\rm tot}(\bar B\rightarrow J/\psi\bar K)
                                                \,(\%) \quad$
& {$\quad$0.075$\quad$}
&{$\quad$0.11$\quad$}
&{$\quad$0.14$\quad$}
\\
$\quad {\cal B}_{\rm tot}(\bar B\rightarrow J/\psi\bar K)
                                                \,(\%) \quad$
& {$\quad$0.075$\quad$}
&{$\quad$0.11$\quad$}
&{$\quad$0.14$\quad$}
\\
\hline
 &\multicolumn{3}{|c|}
{}\vspace{-4mm}\\
{$\quad$Experiments$\quad$}
&\multicolumn{3}{|c|}
{\begin{tabular}{c}
$\hspace{3mm}{\cal B}(B^-\rightarrow J/\psi K^-) 
= (0.100 \pm 0.010)\,\,\%\hspace{3mm}$ 
\\
$\hspace{3mm}{\cal B}(\bar B^0\, \rightarrow J/\psi\bar K^0\,) 
= (0.089 \pm 0.012)\,\,\%\hspace{3mm}$
\end{tabular}
}
\\
\hline\hline
\end{tabular}
\end{center}
\vspace{0.5 cm}
$V_{cs} \simeq V_{ud} \simeq 0.97$. 
The form factor $F_1^{(\bar K\bar B)}(m_\psi^2)$ has not been 
measured and its theoretical estimates are model dependent. We pick 
out tentatively three typical values of 
$F_1^{(\bar K\bar B)}(m_\psi^2)$ 
by the models, BSW\cite{BSW}, 
CDDFGN\cite{CDDFGN} and GKP\cite{GKP}, among many models and list the 
resulting ${\cal B}_{\rm FA}(\bar B \rightarrow J/\psi\bar K)$ in 
Table~4, where we have used the same values of parameters as before. 
${\cal B}_{\rm FA}$ from the factorized amplitude for 
the value of $F_1^{(\bar K\bar B)}(m_\psi^2)$ by BSW, which is close 
to the most recent value from pQCD~\cite{pQCD} with $SU_f(3)$ 
symmetry, is about a half of the observations~\cite{PDG00}, 
${\cal B}(B^- \rightarrow J/\psi K^-)_{\rm exp} 
= (0.100 \pm 0.010) \,\,\%$ and 
${\cal B}(\bar B^0 \,\rightarrow J/\psi\bar K^0\,)_{\rm exp} 
= (0.089 \pm 0.012) \,\,\% $. 
It means that some other (non-factorizable) contribution is needed in
this case. On the other hand, ${\cal B}_{\rm FA}$'s for the higher
values (CDDFGN and GKP) of the form factor saturate the above data
values even if non-factorizable contributions are not included. It
will be seen later that the predicted values of ${\cal B}_{\rm tot}$ 
including non-factorizable contributions is a little too large, in
particular, in the case of GKP. 

Non-factorizable contributions to these decays are estimated by using 
a hard kaon approximation which is a simple extension of the hard 
pion technique in the previous section. With this approximation and
isospin symmetry, non-factorizable amplitude for the 
$\bar B \rightarrow J/\psi\bar K$ decays is given by 
\begin{equation}
M_{\rm NF}(\bar B \rightarrow J/\psi\bar K) 
= {i \over f_K}
{\langle \psi|H_w|\bar B_s^{*0} \rangle }
\Biggl({m_B^2 - m_\psi^2 \over m_{B_s^*}^2 - m_\psi^2}\Biggr)
\sqrt{1 \over 2}h \, +\, \cdots, 
\end{equation}
where the ellipsis denotes neglected contributions of excited 
mesons\cite{Close}. Here we have used 
$\langle{\bar B_s^0|V_{K^+}|B^-}\rangle = -1$ and 
$\sqrt{2}\langle{\bar B_s^{*0}|A_{K^+}|B^-}\rangle = -h$ 
which are flavor $SU_f(3)$ extensions of Eqs.(\ref{eq:MEV}) and 
(\ref{eq:MEA}). 
Asymptotic matrix element, 
$\langle{\psi|\tilde H_w|\bar B_s^{*0}}\rangle$, 
is parameterized in a way similar to that of 
$\langle{D^{*0}|\tilde H_w|\bar B^{*0}}\rangle$, 
i.e., 
$\langle{\psi|\tilde H_w|\bar B_s^{*0}}\rangle 
= B_H'|\langle{\psi|H_w^{\rm BSW}|\bar B_s^{*0}}\rangle_{\rm FA}|$,  
where $B_H'$ is a parameter corresponding to $B_H$ in 
Eq.~(\ref{eq:AME-Hw}). Then the total amplitude for the 
$\bar B \rightarrow J/\psi\bar K$ decays is approximately given by 
\begin{equation}
M_{\rm tot}(\bar B \rightarrow J/\psi\bar K) 
\simeq -iV_{cb}V_{cs}\bigl\{6.12F_1^{(\bar K\bar B)}(m_\psi^2)\,
+\,5.27B_H'\bigr\}a_2\times10^{-5}\,\, {\rm GeV},         
                                                 \label{eq:amp-psi-k}
\end{equation}
where $f_K\simeq 160$~MeV and $f_{B^*_s}\simeq f_{B_s}\simeq 204$~MeV 
from the updated lattice QCD result\cite{Lattice} have been taken. 

When we take $a_2=0.21$ as before and assume tentatively 
$B_H'= B_H$ ($=0.17$ as before), our result, ${\cal B}_{\rm tot}$, 
for the smaller value of the form factor, 
$F_1^{(\bar K\bar B)}(m_\psi^2)$, from BSW [or pQCD with $SU_f(3)$ 
symmetry] is improved considerably. Contrary, ${\cal B}_{\rm tot}$ 
for the larger values of the form factor, in particular, from GKP is 
beyond the measured ones as seen in Table~4. 

For the CKM-angle suppressed $\bar B \rightarrow J/\psi\pi$, 
the same technique and values of parameters as the above lead to 
\begin{eqnarray}
&& M_{\rm tot}(B^- \rightarrow J/\psi\pi^-) 
\simeq -\sqrt{2}M_{\rm tot}(\bar B^0 \rightarrow J/\psi\pi^0) 
                                                        \nonumber\\
&&\simeq -iV_{cb}V_{cd}\bigl\{6.12F_1^{(\pi\bar B)}(m_\psi^2)\,
                +\,6.64B_H'\bigr\}a_2\times10^{-5}\,\, {\rm GeV}.      
                                              \label{eq:amp-psi-pi}
\end{eqnarray}
The first equation is consistent with the measurements, 
${\cal B}(B^-\rightarrow J/\psi\pi^-)_{\rm exp} 
= (5.1 \pm 1.5)\times 10^{-5}$~\cite{PDG00} and 
${\cal B}(\bar B^0\rightarrow J/\psi\pi^0) 
= (2.5^{+1.1}_{-0.9}\pm 0.2)\times 10^{-5}$~\cite{Avery}.
Using $F_1^{(\pi\bar B)}(m_\psi^2)\simeq 0.56$ estimated by 
pQCD~\cite{pQCD}, which is close to the one of 
$F_1^{(\bar K\bar B)}(m_\psi^2)$ from BSW, we obtain 
\begin{equation}
{\cal B}_{\rm tot}(B^- \rightarrow J/\psi\pi^-) 
\simeq 4.2\times10^{-5}, 
\end{equation}
which should be compared with the the above measurement. 

As seen in Table~4, our result on the color suppressed 
$\bar B\rightarrow J/\psi\bar K$ and $\bar B\rightarrow J/\psi\pi$ 
decays is still sensitive to the values of both of 
$F_1^{(\pi\bar B)}(m_\psi^2)$ [or $F_1^{(\bar K\bar B)}(m_\psi^2)$] 
and $B_H'$. It implies that both of the factorized and 
non-factorizable amplitudes are important in these decays. 

\section{Summary}

In the previous sections, we have studied the 
$\bar B \rightarrow D\pi$, $D^*\pi$, $J/\psi \bar K$ and $J/\psi\pi$ 
decays describing their amplitude by a sum of factorizable and 
non-factorizable ones. The former has been estimated by using the 
naive factorization while the latter has been calculated by using a 
hard pion (or kaon) approximation in the infinite momentum frame. 
The so-called final state interactions (corresponding to non-leading 
terms in the large $N_c$ expansion) have been included in the 
non-factorizable long distance contributions. The corresponding ones 
in the color favored $\bar B \rightarrow D\pi$ and $D^*\pi$ decays 
are rather small and therefore the final state interactions seem to 
be not very important in these decays although still not necessarily 
negligible. Next, we have investigated phenomenologically the 
$\bar B \rightarrow D\pi$ and $D^*\pi$ decays, and observed that the 
existing data on their branching ratios are not always compatible 
with each other, i.e., $\cos(\delta^{[*]})$ is over unity for some 
values of $R_{00}^{[*]}$ and $R_{0-}^{[*]}$, where $\delta^{[*]}$ 
denote phase differences between amplitudes for the decays into 
$I={1\over 2}$ and ${3\over 2}$ $D^{[*]}\pi$ final states. 
Eliminating such values of $R_{00}^{[*]}$ and $R_{0-}^{[*]}$, we have 
obtained phenomenologically allowed values of branching ratios, 
${\cal B}_{\rm ph}$, in Table~3, which keep approximately 
$\cos(\delta^{[*]}) < 1$ and include lower limits on 
${\cal B}(\bar B \rightarrow D^0\pi^0)$ and 
${\cal B}(\bar B \rightarrow D^{*0}\pi^0)$. 

By taking $a_1\simeq 1.024$ with the LO QCD corrections and the 
phenomenological $a_2\simeq 0.21$ which has been suggested 
previously~\cite{Kamal,CLEO}, the observed branching ratios for these 
decays can be well reproduced in terms of a sum of the hard pion 
amplitude and the factorized one. Namely, the factorized amplitudes 
are dominant but not necessarily complete and long distance hadron 
dynamics should be carefully taken into account in hadronic weak 
interactions of $B$ mesons. In color suppressed 
$\bar B^0 \rightarrow  D^0\pi^0$, $D^{*0}\pi^0$, 
$\bar B\rightarrow J/\psi\bar K$ and $J/\psi\pi^-$ decays, 
non-factorizable long distance contributions are more important. 
In particular, in the $\bar B \rightarrow J/\psi\bar K$ decays, long 
distance physics should be treated carefully. When $a_2\simeq 0.125$ 
with the LO QCD corrections is taken instead of the phenomenological 
$a_2\simeq 0.21$, it is hard to reproduce the measured 
${\cal B}(\bar B \rightarrow J/\psi \bar K)_{\rm exp}$ and 
${\cal B}(B^- \rightarrow J/\psi\pi^-)_{\rm exp}$ 
even if a sum of factorized and non-factorizable amplitudes is taken 
as long as $B_H'=B_H\simeq 0.17$. 

The non-factorizable amplitudes are proportional to asymptotic 
ground-state-meson matrix elements of $\tilde H_w$, i.e., $B_H$ or 
$B_H'$. To reproduce the measured rates for the color favored 
$\bar B \rightarrow D\pi$ and $D^*\pi$ decays, the non-factorizable
contributions may not be negligible ($B_H \neq 0$) while too large 
values of $B_H$ and $B_H'$ will lead to too large rates for the color 
suppressed decays. However, their numerical results are still 
ambiguous since the amplitudes for the color suppressed decays 
are sensitive to model dependent form factors. 

Experimental data on exclusive decays, in particular, decays into
charm-less final states, of $B$ mesons are rapidly increasing after 
the $B$ factories. In these decays, the factorization will not 
be a good approximation since the large $N_c$ argument does not work 
in hadronic weak processes and since these decays are outside of 
applicability of the color transparency, so that non-factorizable 
contributions are expected to be very important in these decays. 
Besides, some of these decays are expected to play important roles 
in determination of $CP$-violating parameters. Therefore, 
non-factorizable contribution in exclusive decays of $B$ mesons is  
now and will be, in near future, one of very important subjects. 

\vspace{5mm}

The author would like to thank Prof. S.~Pakvasa, Prof. H.~Yamamoto, 
Prof. S.~F.~Tuan and the other members of high energy physics group, 
University of Hawaii for their discussions, comments and hospitality 
during his stay there, and Dr. D.-X.~Zhang for comments. 
He also would like to appreciate Prof. T. Kurimoto for sending 
numerical values of form factors by pQCD.


\end{document}